Research Article

# Trust, Usefulness, and Dependency on AI in Programming: A Hierarchical Clustering Approach

Hilene E. Hernanez[1], Ranie B. Canlas[1], Madilaine Claire B. Nacianceno[1], Jordan L. Salenga[1], Jaymark A. Yambao[1], Juvy C. Grume[1], Aileen P. De Leon[1], Freneil R. Pampo[1], John Paul P. Miranda[1*]

1. **Don Honorio Ventura State University**, Pampanga, Philippines

**\* Correspondence:**
John Paul P. Miranda, Don Honorio Ventura State University, jppmiranda@dhvsu.edu.ph

## ABSTRACT

While AI tools are transforming programming education, their adoption in underrepresented countries remains insufficiently studied. Understanding students' trust, perceived usefulness, and dependency on AI tools is essential to improving their integration into education.  For these purposes, this study surveyed 508 first-year programming students in Pampanga, Philippines and analyzed their perceptions using hierarchical clustering. Results showed four unique student profiles with varying in trust and usage intensity. While students acknowledged AI tools' benefits, dependency remained low due to limited infrastructure and insufficient exposure. High-frequency users did not necessarily report greater trust or usefulness which may indicates a complex relationship between usage patterns and perception. This study recommends that to maximize AI's educational impact, targeted interventions such as infrastructure development, training programs, and curriculum integration are necessary. This study provides empirical insights to support equitable and effective AI adoption in programming education within developing regions.

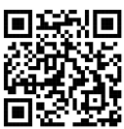

# INTRODUCTION

The increasing integration of Artificial Intelligence (AI) in education has transformed instructional methods and student interactions with technology. For novice programmers, AI tools have become crucial resources for mastering complex concepts and developing skills (Skripchuk et al., 2024; Zviel-Girshin, 2024). However, student perceptions of these tools' usefulness and trustworthiness vary which may influence their potential dependency on AI tools (Robinos et al., 2024; Skripchuk et al., 2024; Zhou et al., 2024). Analyzing student profiles based on perceived usefulness and trust in AI is essential for understanding engagement patterns and optimizing AI-enhanced educational tools in introductory programming courses.

This study examines the profiles of AI dependency among Filipino first-year programming students based on their profile, perceived usefulness, and trust in AI tools. Furthermore, it investigates whether significant differences exist in AI dependency across these identified profiles. Through analyzing these relationships, the research expands current understanding of AI dependency patterns and contributes to emerging discourse on student engagement with AI tools in programming education particularly in underrepresented contexts. The results could advance theoretical frameworks on technology acceptance and trust while providing empirical evidence for developing targeted educational interventions.

# METHOD

The study employed a purposive sampling strategy to recruit first-year programming students (subsequently referred as students) from Pampanga, Philippines. This technique was chosen to ensure that participants had relevant experience with AI tools in academic contexts. Students were selected based on specific inclusion criteria: a prior or current experience using AI tools for academic tasks in their introductory programming courses. This approach allowed for a targeted selection of individuals most likely to provide insights into the research questions.

Students were drawn from both public and private tertiary institutions in October to November 2024. Informed consent was obtained from all participants before data collection. A total of 508 students participated in the study during the midterm period of their classes. The sample comprised 74.01% male (n = 376), 22.83% female students (n = 116), and 2.56% (n = 13) who chose not to disclose their sex assigned at birth. The students age ranged from 17 to 34 years with a mean age of 18.94 years (SD = 1.85). Students reported spending between 1 and 50 hours per week using AI tools for programming-related academic tasks, with an average weekly usage of 4.80 hours (SD = 5.87).

The main instrument in this study measured the perceived usefulness (3 items), trust (5 items), and AI dependency (2 items). The items for perceived usefulness were based from previous studies that discussed the capability of AI tools in enhancing programming efficiency, accelerate task completion, and help students achieve their learning goals by obtaining instant feedback, suggestions, and assistance in programming (Boguslawski et al., 2024; Pan et al., 2024; Sun et al., 2024; Zhong et al., 2024). For this purpose, perceived usefulness in this study is about the extent to which students believe AI tools can enhance their performance, efficiency, or learning outcomes in the context of programming. It is believed that as student experience the benefits of AI tools, they tend to rely more on these tools. Over time, this can evolve into AI dependency where students view AI tools as an essential tool for completing tasks they might otherwise struggle with.

In AI trust, the items stems from the reliability, accuracy, and ability of AI tools in providing actionable insights and advice in relation to specific tasks particularly in the academic contexts (Afroogh et al., 2024; Boguslawski et al., 2024). AI trust in this study refers to the confidence of the students in AI tools in its ability to provide reliable, accurate, and helpful suggestions or outputs. In relation to AI dependency, trust is crucial as it can reduce hesitation and can increase the likelihood of students to adopt and rely more on AI tools. This shows that perceived usefulness can drive adoption of these AI tools because of its value while trust ensures sustained dependency by building confidence in their ability. These indicates that both perceived usefulness and trust creates a strong foundation and shape the relationship for AI dependency (Jeong et al., 2025). For AI dependency, the items were derived from studies that show AI tools is becoming an integral part in programming education particularly due to their convenience, accessibility, and ability to simplify complex tasks (Boguslawski et al., 2024; Sun et al., 2024; Zhong et al., 2024). Despite its brevity, the construct was conceptually aligned to capture students' dependency on AI tools in programming through two key dimensions: importance and frequency of use. All items utilized a 5-point Likert scale with some items are scored in reverse. These were ranged from "strongly disagree" to "strongly agree." All constructs were validated by IT professors and passed the required Cronbach's alpha (i.e., > 0.70).

In reporting the results, descriptive statistics were described and hierarchal clustering was performed to identify distinct profiles based on perceived usefulness and AI trust. Hierarchical clustering was selected because it reveals nested data structures which enhances the interpretability of similarity patterns within a dataset (Boyko & Tkachyk, 2023; Ran et al., 2023). Unlike K-means or other similar techniques, it does not require pre-specified cluster numbers, provides dendrogram visualization, and demonstrates greater robustness for small to medium datasets which makes it particularly suitable for this study (Boyko & Tkachyk, 2023). Furthermore, to strengthen the use of hierarchal clustering, Silhouette Score and Elbow Method were used to support the optimal number of clusters. The first method evaluated the within-cluster sum of squares (WCSS) to identify the point where additional clusters yielded diminishing returns in variance reduction. While the latter quantified cluster cohesion and separation with higher values that shows well-delineated clusters. Euclidean distance served as the standard metric for the assessment of inter-point similarity which helped in validating the clusters. The characteristics of each profile were then examined and differences in AI dependency across the identified profiles were tested. All analyses were carried out using SPSS version 25 and Python programming within the Jupyter Notebook environment.

# RESULTS AND DISCUSSION

## Descriptive Results

Table 1 reveals that students reported slight agreement regarding the usefulness of AI tools for programming tasks. This suggests that while they recognize the value of these tools, their perception is tempered by certain limitations (Akçapınar & Sidan, 2024; Boguslawski et al., 2024; Jose et al., 2022; Liang et al., 2024; Zviel-Girshin, 2024). Similarly, students express moderate agreement on statements related to trust in AI tools. This indicates that while they do trust these tools to some extent, their confidence is not particularly strong, likely due to concerns about the tools' lack of guidelines, reliability, explainability, and potential biases (Labadze et al., 2023; Saari et al., 2024). In terms of dependency, students report that they do not feel highly dependent on AI tools. This is likely attributable to the limited integration of such tools in educational settings, particularly in developing countries like the Philippines, where the study was conducted (Akçapınar & Sidan, 2024; Jose et al., 2022; Saari et al., 2024; Zviel-Girshin, 2024). Factors such as inadequate infrastructure, limited access and exposure to advanced technology, and insufficient training for both students and educators may contribute to this low level of dependency.

**Table 2.**

*Descriptive statistics*

| Items | Mean | SD |
|---|---|---|
| **Perceived Usefulness** | | |
| AI tools help me complete programming tasks faster. | 3.33 | 0.905 |
| AI tools make it easier for me to reach my learning goals in programming. | 3.41 | 0.874 |

| Items | Mean | SD |
|---|---|---|
| **Perceived Usefulness** | | |
| Using AI tools makes my programming work more effective. | 3.32 | 0.926 |
| **AI Trust** | | |
| I trust AI tools to give me useful advice for programming. | 3.21 | 0.868 |
| AI tools give helpful ideas for making good programming decisions. | 3.33 | 0.865 |
| I feel confident using suggestions from AI tools in my programming work. | 3.18 | 0.839 |
| I plan to use AI tools even more in my programming work. | 3.23 | 0.875 |
| I rely on AI tools to complete programming tasks I might struggle with on my own. | 3.17 | 0.818 |
| **AI Dependency** | | |
| AI tools are an important part of how I do programming. | 3.15 | 0.799 |
| I often use AI tools to help me with programming tasks. | 3.16 | 0.865 |

## Hierarchal Clustering

Figure 1a shows the Elbow Method indicated 4 clusters as optimal where the WCSS stabilizes. The Silhouette Score in Figure 1b reaches its maximum at 2 clusters and decreases thereafter. The score remains moderate at 4 clusters which suggests well-defined groups while 5 clusters exhibit reduced cohesion. These metrics confirm that 4 clusters offer the optimal balance between compactness and interpretability of the dataset for hierarchical clustering.

**Figure 1.**
*Elbow method (a) and Silhouette score (b)*

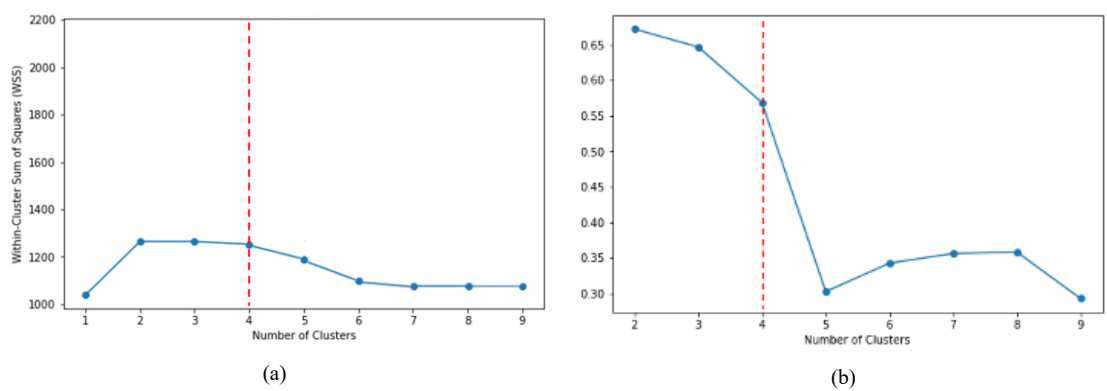

(a)        (b)

The four clusters identified through hierarchal clustering (Fig. 2) were subsequently referred as profile and represent distinct characteristics based on their PU, AT, and Hours spent, alongside demographic variables such as Age and Gender (Table 2). For instance, Profile 1 comprises users with moderate levels of both PU and AT. These users spend minimal time using AI tools, averaging 3.25 hours per week with an average age of 18.96, and the majority (75%) are male. This may be explained by their limited exposure to AI (Gerlich, 2023; Grassini, 2023; Klarin et al., 2024; Robinos et al., 2024). Although one study has indicated that even with limited exposure to such tools, positive attitudes can also be seen if these tools are intuitive and provide immediate value to them. For instance, among younger users, AI-driven educational tools have been shown to enhance perceived usefulness even with minimal engagement (Burkhard, 2022; Fošner, 2024; Klarin et al., 2024; Lijie et al., 2024).

Profile 2, in contrast, represents the heaviest users of AI tools, spending an average of 48.5 hours per week. Despite their high usage, they exhibit the lowest levels of PU and AT. They have an average age of 18.25 and consists exclusively of males. This paradox mirrors earlier findings where extensive use of AI tools does not necessarily correlate with higher PU or AT, possibly due to unmet expectations or challenges in integration (Grassini, 2023; Lijie et al., 2024; Yang & Wibowo, 2022). This is also contrary to some study that suggest users who invest more time in such tool are more likely to discover their benefits (Doe et al., 2017; Gao & Wang, 2024). One possible explanation to this contradiction is the setting and demographics of the study (i.e., Philippines, a developing country) (Cirera et al., 2022).

**Figure 2.**
*Dendrogram of hierarchical clustering with four clusters*

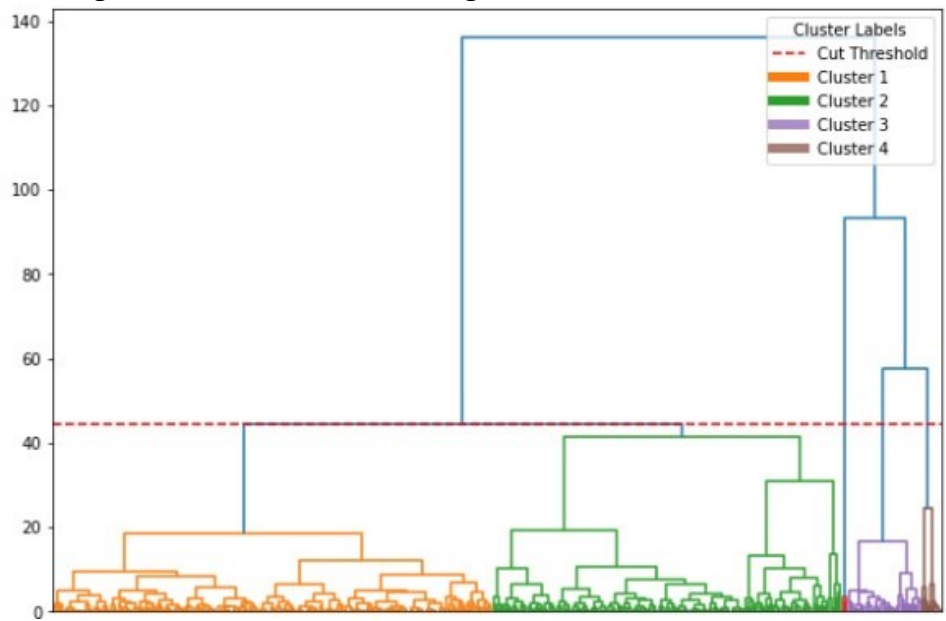

Profile 3 is characterized by the highest levels of PU and AT. Members of this profile demonstrate moderate engagement with AI tools, using them for an average of 11.36 hours per week. They have an average age of 18.86, with 67% being male. This suggests that balanced engagement may allow users to appreciate AI benefits without encountering significant drawbacks (e.g., cognitive overload or frustration) and potentially support existing research that moderate use can foster positive attitudes (Faruqe et al., 2023; Gerlich, 2025; Jeong et al., 2025; Lijie et al., 2024; Robinos et al., 2024). However, factors such as the quality of AI tools, training, and contextual relevance may play an important role in shaping and sustaining such attitudes (Geddam et al., 2024; Grassini, 2023; Jo & Bang, 2023).

Profile 4 also shows high levels of PU and AT but is distinguished by higher engagement, with users spending an average of 24.67 hours per week on AI tools. This profile has an average age of 18.58 with 75% of its members being male. This indicates that users who find AI tools useful and trustworthy are inclined to integrate them more into their routines. This result is aligned with earlier studies on the positive feedback loop between PU and usage intensity (Jeong et al., 2025; Kang et al., 2024; Pan et al., 2024). However, it was cautioned by experts that excessive dependency on AI tools can lead to over-dependence and may result to reduce thinking and creativity and that the relationship between usage intensity and positive attitudes may have diminishing returns (Jackson, 2025; Knapp, 2025; Zhai et al., 2024).

The results reveal interesting trends across the four profiles. Profile 2 demonstrates a notable contrast between their high engagement with AI tools and their lower levels of PU and AT, which may indicate dissatisfaction or unmet expectations. On the other hand, Profile 3 and 4 are characterized by high levels of PU and AT, with Profile 4 users spending significantly more time on AI tools than Profile 3. This suggests that Profile 4 users may perceive greater utility or value from these tools, which could explain their higher engagement levels. Profile 1, with its moderate levels of PU and AT coupled with minimal usage, may represent casual or novice users who are less reliant on AI tools for their programming tasks.

**Table 2.**

*Profile summary*

| Cluster | PU (Mean ± SD) | AT (Mean ± SD) | Hours (Mean ± SD) | Age (Mean ± SD) | Gender |
|---|---|---|---|---|---|
| 1 | 3.35 ± 0.79 | 3.21 ± 0.68 | 3.25 ± 1.93 | 18.96 ± 1.92 | 0.75 |
| 2 | 2.75 ± 0.88 | 2.95 ± 0.98 | 48.5 ± 1 | 18.25 ± 0.5 | 1.00 |
| 3 | 3.54 ± 0.68 | 3.43 ± 0.65 | 11.36 ± 2 | 18.86 ± 1.28 | 0.67 |
| 4 | 3.36 ± 0.52 | 3.25 ± 0.4 | 24.67 ± 5.57 | 18.58 ± 0.9 | 0.75 |

### Results of AI Dependency in Each Profile

A Kruskal-Wallis H test revealed no statistically significant differences in AI tools dependency across the four identified profiles, $\chi^2(3)$ = 3.297, p = 0.348. The mean ranks of AI dependency were as follows: Profile 1 (253.83), Profile 2 (166.00), Profile 3 (278.49), and Profile 4 (225.21). These results suggest that AI dependency levels are comparable among the profiles identified through hierarchical clustering. This indicates that the observed uniformity in AI dependency may be attributed to the limited engagement with AI tools previously reported in the student demographics. As first-year programming students, this is probably due to their limited exposure to programming concepts and AI tools which could account for the lack of significant variation in dependency. This inexperience coupled with not so wide standard deviation with AI tool usage may suggests that AI tools are supplementary rather than central to students' learning practices at this stage. These factors may explain why no meaningful differences in AI dependency emerged across the profiles. Furthermore, the consistently low AI dependency in this study may be specific to the Philippines. Schools in the country face significant technological and infrastructural limitations (Fabito et al., 2021). Specifically, public universities often struggle with inconsistent internet access, a lack of necessary devices, limited subscriptions to online resources, and inadequate educator training in digital technologies including AI (Kunjiapu et al., 2024). These factors pose substantial barriers to technology adoption Consequently, the absence of a national or uniform policy on AI integration in higher education curricula likely contributes to the low dependency. Additionally, some studies suggests that both teachers and students perceived the use of AI for academic work as unethical which might further explain the observed trend (Cavazos et al., 2024; Jason V. Chavez et al., 2024).

## CONCLUSION AND RECOMMENDATIONS

The study highlights that while students generally acknowledge the usefulness and trustworthiness of AI tools in programming tasks, their perceptions remain moderate. These perceptions are shaped by contextual limitations including insufficient infrastructure, inadequate training, and limited exposure. Hierarchical clustering analysis identified four distinct user profiles. The study recommends that it is crucial to address key barriers such as limited infrastructure, exposure, and insufficient training. Introducing targeted seminars, workshops, and incorporating AI tool training into the curriculum could improve familiarity and confidence to help student better understand the potential benefits of AI tools given that such skill is increasingly important. For instance, students can leverage tools like GitHub Copilot or ChatGPT to analyze and explain coding errors in real-world scenarios before manual debugging. This fosters critical error analysis skills while mitigating over-reliance on AI-generated solutions. Furthermore, incorporating AI-powered code analysis tools that assess efficiency, reliability, and correctness can significantly strengthen students' programming logic. The use of AI-assisted platforms and GPT-based tutors may also be used to generate adaptive coding challenges that suit the skill levels of individual students.

For future works, this study recommends that researchers should also explore changes in students' perceptions and usage patterns over time in the context of underrepresented countries the Philippines. For instance, longitudinal studies may provide deeper insights into how AI dependency evolves as students gain more programming experience and whether increased exposure leads to greater reliance on AI tools. Additionally, future research can examine how the user profiles identified in this study align with or challenge existing theoretical frameworks, such as the Technology Acceptance Model (TAM). Investigating factors like perceived ease of use, perceived usefulness, and behavioral intention within these clusters can offer a more structured understanding of AI adoption in programming education.